\documentstyle[12pt]{article}
\newcommand{\be}{\begin{equation}}
\newcommand{\ee}{\end{equation}}
\newcommand{\bea}{\begin{eqnarray}}
\newcommand{\eea}{\end{eqnarray}}
\newcommand{\ba}{\begin{array}}
\newcommand{\ea}{\end{array}}
\newcommand{\beas}{\begin{eqnarray*}}
\newcommand{\eeas}{\end{eqnarray*}}
\newcommand{\bes}{\begin{equation*}}
\newcommand{\ees}{\end{equation*}}

\begin{document}
\title{\bf The stress energy tensor of neutral blackfold and dual theory}
\author{ Z. Amoozad  \thanks{E-mail: z.amoozad@stu.umz.ac.ir}
         and           J. Sadeghi   \thanks{E-mail: pouriya@ipm.ir }\hspace{1.5mm}\\
{\small {\em Sciences Faculty, Department of Physics, University of Mazandaran,}}\\
{\small {\em  Babolsar, Iran, P.O.Box 47416-95447 }}\\
} \maketitle

\begin{abstract}
 In this paper we consider charged and neutral blackfold and extract the Brown-York stress energy tensor. Also, we show that the neutral blackfold spacetime is Ricci-flat and the other spacetime is not. This Ricci-flat condition gives us opportunity to calculate the AAdS spacetime. In order to have dual theory one can consider the AAdS in Fefferman-Graham coordinates. This frame gives correct form of stress tensor in the boundary. The corresponding tensor with using this frame will be traceless and conserved. Such stress tensor is same as perfect fluid and it proves the dual renormalized theory exists for the neutral blackfold . 
\end{abstract}

\section{Introduction}
\label{intro}
One of the best subjects which is originated from the study of quantum gravity and string theory is holographic principle which is proposed by Gerard 't Hooft in 1993 ~\cite{a,b,c,d,e}. He has shown that physics in region of space with gravity could be regain in dual space which does not contain gravity and lives on a reduced space with smaller dimension. Although the reduced space is a hypersurface of the original background ~\cite{z}. Typically it relates gravity to gauge theory which gives an equivalence between D dimensional gravitational theory and D-1 dimensional quantum field theory, which is known as gauge/gravity duality.

 The most important gauge theory/gravity example is AdS/CFT which is given by Maldacena. It connects asymptotically AdS(AAdS) geometry to the states in the field theory. For example the pure AdS bulk geometry corresponds to the vacuum state and deformed  bulk geometry in AdS side corresponds to an exited state in the gauge theory side.

 AdS/CFT correspondence has been applied to various phenomena on various scale such as QCD, condensed matter, cosmology and black holes. In this paper we work on black hole as a candidate for the AdS side. An interesting aspect of this duality exhibits when we go to long wave length regime. In this regime quantum field theory describes the existence of hydrodynamics and fluid/gravity duality could be introduced ~\cite{f,g,h}.

Also many different subjects have been investigated in these grounds. For example, AdS black brane solutions correspond to conformal and non conformal fluid at thermal equilibrium ~\cite{l,m} and also some relativistic expansion of fluid/gravity duality have been studied in  ~\cite{n,o}. Extension of fluid/gravity duality to spherical horizon topologies, de sitter spacetime and charged fluid are calculated in ~\cite{p,q,r,s,t,u,v}.

All above discussion  may exhibit new results if we extend 4-dimensional spacetimes to higher dimensional ones. But developing relations and calculations to higher dimensions requires some new tools and methods, because of the different properties of higher dimensional black holes.  By definition, if black holes regarded as a black brane (possibly boosted locally) which folded into multiple dimensions a blackfold could be shaped. One of the main difference of higher dimensional black hole with respect to the 4-dimensional ones is that the horizons of higher dimensional black holes can have at least two different size lengths. One reason is that they haven't any bound on their angular momentum, although in 4-dimension the Kerr bound $J \leq GM^{2}$ does not admit higher momentum. So in blackfolds, there is two length scales which  are given by,

\begin{equation}
\textit{l}_M \sim (GM)^{1/D-3}\;\;\;\;\  ,  \;\;\;\;\;\;\ \textit{l}_J \sim \frac{J}{M}
\end{equation}

 In case of $\textit{l}_J \gg \textit{l}_M $ , this separation suggests an effective description of long wavelength dynamics~\cite{i,j,k}.

   At the other hand there are black p-branes that carry charges of Ramond-Ramond field strength $F_{(p+2)}$ and play an important role in string theory. They are very significant in higher-dimensional supergravity theories, especially in low energy limit of string/M theory ~\cite{aaa}.

 All above information from black hole give us motivation to investigate two candidates, charged and neutral blackfolds. So we take advantages from stress tensor in black hole family and find some dual theory. We have seen that the charged blackfold is not compatible with dual theory because the corresponding tensor could not be traceless even by renormalization. Also the charged blackfold solution is not Ricci-flat, so we have not AAdS form.

In detail we use Brown-York method and extract quasilocal stress tensor. It could not exactly be boundary stress energy tensor of fluid because it calculates on the hypersurface which is not on the boundary and just could be related to that in some special cases ~\cite{h}. For this reason the attached stress tensor could not be considered as the dual stress tensor. To find dual stress tenor we use the AdS/Ricci-flat correspondence and neutral blackfold as a Ricci-flat case, then we extract AAdS form of a neutral blackfold in the Fefferman-Graham coordinate. Then the dual renormalized stress tensor which is a property of the boundary would be obtained~\cite{ ee,ff,aa,bb,cc}. For the case of charged blackfold, as it has not the constraint of Ricci-flat, the AdS form of that could not be exist and so it does not have dual description.

To obtain our purpose this paper is organized as follows. In the next section we introduce charged and neutral blackfolds and discuss some of their properties and present their differences and similarities. In section 3 by using background metrics for charged and neutral blackfolds explicit expressions for Brown-York stress energy tensor on a time-like hypersurface will be extracted. In section 4 by checking the required condition, and applying AdS/Ricci-flat correspondence, we obtain dual renormalized holographic stress tensor for neutral blackfold. In final section we conclude and note some points.

 \section{Preliminary: Geometry of charged and neutral blackfold}
\label{sec:1}

The 4-dimensional black hole has only short scale dynamics with comparing to higher dimensional black branes which has two different size horizons ($\sim R \gg r_0 $). This property captures the long distance dynamics, so in that case there is an effective worldvolume theory. Therefore for higher dimensional black branes new approach and methods are needed to work. When the worldvolume of a black p-brane bents and formed as a compact hypersurface, the extracted black brane is named $blackfold$.

In supergravity and low energy limit of string theory, black branes have charges. One of the best choices of charged black p-branes is Ramond-Ramond field strength as $F_{(p+2)}$. The action of charged dilatonic black p-brane in $D=n+p+3$ spacetime dimension is ~\cite{y},

\begin{equation}
\label{eq:x}
I=\frac{1}{16\pi G}\int dx^{D} \sqrt{-g}(R-\frac{1}{2}(\partial \phi)^{2}-\frac{1}{2(p+2)!}e^{a\phi} F^{2}_{(p+2)}),
\end{equation}
where
\begin{equation}\label{eq:x}
a^{2}=\frac{4}{N}-\frac{2(p+1)n}{D-2}  \;\;\;\;\;\;  ,  \;\;\;\;\;\;    n=D-p-3.
\end{equation}

Gravitational description of the world volume theory would be appropriate when we have a stack of N of p-branes. The metric of charged flat black p-brane is,

\begin{equation}
\label{eq:x}
ds^{2}=H^{\frac{-Nn}{D-2}}(-fdt^{2}+{\sum _{i=1}}^{p}{dz_{i}}^{2})+H^{\frac{N(p+1)}{D-2}}(f^{-1}dr^{2} +r^{2}{d\Omega ^{2}}_{n+1}),
\end{equation}

\begin{equation}\label{eq:x}
e ^{2\phi}=H^{aN}   \;\;\;,\;\;\; A_{p+1}=\sqrt{N}\coth \alpha (H^{-1}-1)dt \wedge dz_{1} \wedge dz_{2} ... \wedge dz_{p},
\end{equation}

\begin{equation}\label{eq:x}
H=1+\frac{{r_{0}}^{n}\sinh ^{2}\alpha}{r^{n}} \;\;\;\;\;\;\;\;\; ,  \;\;\;\;\;\;\;\;\;    f=1-\frac{{r_{0}}^{n}}{r^{n}},\;\;\;\;\;\;\;\;\;\;\;\;\;
\end{equation}

where $\sinh \alpha$ is related to the boost of the blackfold which is a Lorentz transformation of some directions of p-brane. As $a^{2}\geq0 $,  the parameter $N$ is not arbitrary and there is,
\begin{equation}
\label{eq:x}
N\leq2(\frac{1}{n}+\frac{1}{p+1}),
\end{equation}

In string/M theory, N is an integer up to 3 (when $p\geq1$)that corresponds to the number of different types of branes in an intersection. In this case the effective stress tensor in any hypersurface could be written as,
\begin{equation}
\label{eq:x}
T^{ab}=\tau s(u^{a}u^{b}-\frac{a}{b}\gamma^{ab})-\Phi _{p} Q_{p}\gamma^{ab}.
\end{equation}
where $\tau$ is temperature and s is entropy and $\Phi _{p}$ is the potential that measured from the difference between the values of $A_{p+1}$ at the horizon at $r\rightarrow \infty$  in (5). As one can see, the relation (8) surprisingly has a brane tension component $-\Phi _{p}Q_{p}$ and a thermal component $\tau s $. It must be noted that, blackfolds with p-brane charges do not have open boundaries; as the charge $Q_{p}$ would not be conserved at boundaries. So it is possible to consider that blackfolds end on another brane that carries appropriate  charge.

 Also we can simplify the spacetime and take neutral flat black brane. Effective theory describes the collective dynamics of a black p-brane where it's geometry in $D=n+p+3$ spacetime dimension is ~\cite{k},

\begin{equation}
\label{eq:x}
 {ds^{2}}_{p-brane}=-(1-\frac{{r_{0}}^{n}}{r^{n}})dt^{2}+\sum _{i=1}^{p}{dz_{i}}^{2}+{(1-\frac{{r_{0}}^{n}}{r^{n}})}^{-1}dr^{2} +r^{2}{d\Omega ^{2}}_{n+1},
\end{equation}
where $\sigma ^{a}=(t,z^{i})$ is related to brane worldvolume. By boosting it along the worldvolume, one can easily find following metric,
\begin{equation}
\label{eq:x}
 {ds^{2}}_{p-brane}=(\eta _{ab}+\frac{{r_{0}}^{n}}{r^{n}} u_{a}u_{b})d\sigma^{a}d\sigma^{b}+{(1-\frac{{r_{0}}^{n}}{r^{n}})}^{-1}dr^{2} +r^{2}{d\Omega ^{2}}_{n+1},
\end{equation}
where $r_{0}$ is the thickness of the horizon. Here $(p+1)$ coordinates are on the worldvolume of the blackfold and $(D-p-1)$ coordinates are transverse directions to the worldvolume.

 If $D_{a}$ be the covariant derivative with respect to the boundary metric $\gamma _{ab}$, the equations of conservation of the quasi local stress tensor will takes the following form,
\begin{equation}
\label{eq:x}
 D_{a}{T_{ab}}^{(quasilocal)}=0
\end{equation}
This is the dynamics of effective fluid that lives on the worldvolume and spanned by the brane. For isotropic worldvolume theory, in case of lowest derivative order, the stress tensor will be the same as the isotropic perfect fluid, which is given by,
\begin{equation}
\label{eq:x}
T^{ab}=(\varepsilon+p)u^{a}u^{b}+p\gamma^{ab}
\end{equation}
where $\varepsilon $ is the energy density and $p$ is the pressure. If it doesn't be a perfect fluid then the stress tensor will have dissipative terms proportional to gradients of $ \ln r_{0}, u^{a}, \gamma ^{ab}$. For stationary spacetimes the effect of dissipative terms vanish.

\section{Brown-York quasilocal stress energy tensor}
\label{sec:2}

In the fluid/gravity duality, essential physical properties of field theory side are mentioned by boundary stress tensor. At the other hand in membrane paradigm, stress energy tensor calculates on the stretched horizon. But another quasilocal stress tensor is proposed by Brown-York, which obtain that on a hypersurface which is induced by bulk geometry. As we know the quasilocal stress energy tensor is introduced by Brown and York ~\cite{x}. It appropriately matches the coupling of the short wavelength and long wavelength degrees of freedom. Thus first candidate is charged blackfold whose bulk metric described by (4). Hypersurfaces can be obtained by constraining independent coordinates of spacetime. The time-like hypersurfaces play important role for causality relations between different coordinates. The hypersurface considered here is defined by:
\begin{equation}
\label{eq:x}
r=r_{c},
\end{equation}

As we know, vectors perpendicular to a time-like hypersurface are space-like and time-like hypersurfaces are connected by such vectors. In that case, the Brown-York stress energy tensor ~\cite{v} is defined by,
\begin{equation}
\label{eq:x}
T_{ab}=2(K\gamma_{ab}-K_{ab}),
\end{equation}
where $\gamma_{ab}$ is the induced metric on the time-like hypersurface and $K_{ab}$ is the extrinsic curvature of the bulk. This can be determined by,
\begin{equation}
\label{eq:x}
K_{\mu\nu}=-n_{\mu}a_{\nu}-n_{\nu;\mu}\; ,
\end{equation}
where $n_{\lambda}$ is the unit normal vector to the hypersurface,
\begin{equation}
\label{eq:x}
n_{\lambda}=\frac{\phi_{,\lambda}}{|g^{\mu\nu}\phi_{,\mu}\phi_{,\nu}|^\frac{1}{2}}\; ,
\end{equation}

\begin{equation}
\label{eq:x}
a_{\nu}=n^{\theta}n_{\nu;\theta}\;.\;\;\;\;\;\;\;
\end{equation}
and $\phi$ in the (16) is the equation of hypersurface. In case $r=r_{c}$, $\phi$ just depends on $r$. The charged dilatonic blackfold from (4) could be rewritten as the following,
 \begin{equation}
\label{eq:x}
{ds^{2}}_{p-brane}=A{dt}^{2}+B{dz_{i}}^{2}+Cdr^{2} +F_{\alpha}{d\psi ^{2}}_{\alpha},
\end{equation}
with
\begin{equation}
\label{eq:x}
A=-(1+\frac{{r_{0}}^{n}\sinh ^{2}\alpha}{r^{n}})^{\frac{-Nn}{D-2}}(1-\frac{{r_{0}}^{n}}{r^{n}}),\;\;\;\;\;\;
B=(1+\frac{{r_{0}}^{n}\sinh ^{2}\alpha}{r^{n}})^{\frac{-Nn}{D-2}},
\end{equation}
and
\begin{equation}
\label{eq:x}
C=(1+\frac{{r_{0}}^{n}\sinh ^{2}\alpha}{r^{n}})^{\frac{N(p+1)}{D-2}}(1-\frac{{r_{0}}^{n}}{r^{n}})^{-1},\;\;\;
F_{\alpha}=F_{\alpha}(r,\psi_{\alpha}).\;\;\;\;\;\;\;\;\;\;\;\;\;\;\;\;
\end{equation}
This metric is diagonal, so the inverse matrix can be found easily. For such a metric it is tried to determine the extrinsic curvature (15). In order to obtain the extrinsic  curvature, one can arrange the $n_{\lambda},\; n^{\theta}$ and $a_{\nu} $ as,
\begin{equation}
\label{eq:x}
n_{\lambda}=\sqrt{C}{\delta_{\lambda}}^{r},\;\;\;\;\;\;
n^{\theta}=\sqrt{C}g^{\theta r},  \;\;\;\;\;\;
a_{\nu}=n^{\theta}(n_{\nu ,\theta}-{\Gamma^{r}}_{\nu \theta}\sqrt {C}). \;\;\;\;\;\;\;\;\;\;\;\;
\end{equation}
So, by applying equation (21) into equation (15) one can obtain the corresponding extrinsic curvature as,
\begin{equation}
\label{eq:x}
K_{\mu\nu}=-{\delta_{\mu}}^{r}{\delta_{\nu}}^{r}\frac{C_{,r}}{2\sqrt{C}}+C\sqrt{C}{\delta_{\mu}}^{r}g^{rr}
{\Gamma^{r}}_{\nu r} -  \frac{C_{,\mu}}{2\sqrt{C}}{\delta_{\nu}}^{r}+ \sqrt{C}{\Gamma^{r}}_{\mu\nu}.
\end{equation}
By calculating the Christoffel symbols the nonvanishing components of extrinsic curvature for that geometry are given by,
\begin{equation}
\label{eq:x}
K_{tt}=-\frac{A_{,r}}{2\sqrt{C}}\;\;\;\;,\;\;\; K_{ii}=-\frac{B_{,r}}{2\sqrt{C}}\;\;\;,\;\;\;K_{{\psi}_{\alpha}{\psi}_{\alpha}}=-\frac{F_{\alpha,r}}{2\sqrt{C}}.
\end{equation}
On the other hand, the induced metric $\gamma _{ab}$ on the time-like hypersurface is,

\begin{equation}
\label{eq:x}
ds^{2}|_{r=r_{c}}=A_{c}{dt}^{2}+B_{c}{dz_{i}}^{2}+F_{\alpha c}{d\psi ^{2}}_{\alpha},
\end{equation}
The trace of the extrinsic tensor can be written by following equation,
\begin{equation}
\label{eq:x}
K=K_{ab}\gamma ^{ab}= \frac{-1}{2\sqrt{C}}\{\frac{A_{,r}}{A_{c}}+p\frac{B_{,r}}{B_{c}}+\sum _{\alpha =1}^{n+1}\frac{F_{\alpha,r}}{F_{\alpha c}}\},
\end{equation}
where $p$ is the number of spatial coordinates of worldvolume of blackfold. So the Brown-York stress energy tensor can easily be constructed.
For the case of D/NS-brane in type II string theory ~\cite{w} we have,
\begin{equation}
\label{eq:x}
D=10\;\;\;\;,\;\;\;N=1\;\;\;,\;\;\; p=0,...,6.
\end{equation}
Following the definition,
\begin{equation}
\label{eq:x}
\textbf{A}=\frac{A_{,r}}{A_{c}}\;\;\;,\;\;\;\textbf{B}=\frac{B_{,r}}{B_{c}}\;\;\;,\;\;\;
\textbf{F}_{\alpha}=\frac{F_{\alpha,r}}{F_{\alpha c}}\;\;\;,\;\;\;
\end{equation}
in case of $p=5$ brane $(n=2)$, the nonvanishing component of Brown-York tensor on $r=r_{c}$ are,
\begin{equation}
\label{eq:x}
T_{tt}=\frac{-1}{\sqrt{C}}\{[\textbf{A}+5\textbf{B}+\textbf{F}_{1}+\textbf{F}_{2}+\textbf{F}_{3}]A_{c}-A_{,r}\}\;\;\,,
\end{equation}

\begin{equation}
\label{eq:x}
T_{ii}=\frac{-1}{\sqrt{C}}\{[\textbf{A}+5\textbf{B}+\textbf{F}_{1}+\textbf{F}_{2}+\textbf{F}_{3}]B_{c}-B_{,r}\}\;\;\,,
\end{equation}

\begin{equation}
\label{eq:x}
T_{\alpha\alpha}=\frac{-1}{\sqrt{C}}\{[\textbf{A}+5\textbf{B}+\textbf{F}_{1}+\textbf{F}_{2}+\textbf{F}_{3}]F_{\alpha c}-F_{\alpha ,r}\}.
\end{equation}
One can easily check that there is a trace anomaly here $T_{\mu}^{\mu} \neq 0$ . In Brown-York method we overtake stress tensor on any hypersurface which could be     induced by background metric. In this situation some appropriate counterterm must be added to the action to cancel some divergences which arise here.

Now we take the  neutral blackfold case and try to extract stress tensor of that on the hypersurface. It has been reviewed in the section 2 and the appropriate metric found to be the equation (9). Here the method is the same as before, and we only present $p=5$ brane which characterized by the metric,
\begin{equation}
\label{eq:x}
ds^{2}=A{dt}^{2}+B{dz_{i}}^{2}+Cdr^{2} +F_{\alpha}{d\psi ^{2}}_{\alpha},
\end{equation}
and
\begin{equation}
\label{eq:x}
A=-(1-\frac{{r_{0}}^{2}}{r^{2}}), \;\;\;\;
B=1, \;\;\;\;
C=(1-\frac{{r_{0}}^{2}}{r^{2}})^{-1},\;\;\;\;
F_{\alpha}=F_{\alpha}(r,\psi_{\alpha}).\;\;\;\;
\end{equation}
The method is straightforward and the nonvanishing components of stress tensor for neutral blackfold  at $r=r_{c}$ is,
\begin{equation}
\label{eq:x}
T_{tt}=\frac{6}{r_{c}}(1-\frac{{r_{0}}^{2}}{r_{c}^{2}})^{\frac{1}{2}}(1+\frac{{r_{0}}^{2}}{r_{c}^{2}}),\;\;\;\;\;\;\;\;\;\;\;\;\;\;\;\;\;\;\;\;\;\;\;\;\;\;\;\;\;\;\;\;\;\;\;\;\;\;\;\;\,
\end{equation}

\begin{equation}
\label{eq:x}
T_{ii}=-(1-\frac{{r_{0}}^{2}}{r_{c}^{2}})^{\frac{1}{2}}\{
\frac{\frac{2{r_{0}}^{2}}{r_{c}^{3}}}{(1-\frac{{r_{0}}^{2}}{{r_{c}}^{2}})}+\frac{6}{r_{c}}\},\;\;\;\;\;\;\;\;\;\;\;\;\;\;\;\;\;\;\;\;\;\;\;\;\;\;\;\;\,
\end{equation}

\begin{equation}
\label{eq:x}
T_{\theta\theta}=-{r_{c}}^{2}(1-\frac{{r_{0}}^{2}}{r_{c}^{2}})^{\frac{1}{2}}
\{\frac{\frac{2{r_{0}}^{2}}{r_{c}^{3}}}{(1-\frac{{r_{0}}^{2}}{{r_{c}}^{2}})}+\frac{4}{r_{c}}\},\;\;\;\;\;\;\;\;\;\;\;\;\;\;\;\;\;\;\;\;\;\;\,
\end{equation}

\begin{equation}
\label{eq:x}
T_{\phi\phi}= \sin^{2}\theta \;T_{\theta\theta},\;\;\;\;\;\;\;\;\;\;\;\;\;\;\;\;\;\;\;\;\;\;\;\;\;\;\;\;\;\;\;\;\;\;\;\;\;\;\;\;\;\;\;\;\;\;\;\;\;\;\;\;\;\;\;\;\;\,
\end{equation}

\begin{equation}
\label{eq:x}
T_{\psi\psi}= \sin^{2}\phi \; T_{\phi\phi}.\;\;\;\;\;\;\;\;\;\;\;\;\;\;\;\;\;\;\;\;\;\;\;\;\;\;\;\;\;\;\;\;\;\;\;\;\;\;\;\;\;\;\;\;\;\;\;\;\;\;\;\;\;\;\;\;\,
\end{equation}
Like the charged blackfold, this stress tensor also have trace anomaly which shows that the corresponding lagrangian is not invariant under conformal transformation. It can be seen easily that this quasilocal stress tensor is also conserved.

Now, we are going to find dual theory for the neutral blackfold. Because, this correspondence spacetime is Ricci-flat. This Ricc-flat properties helps us to find AAdS spacetime for the neutral blackfold. Also, note that in the case of charged blackfold there is no form of AAdS because it is not Ricci-flat.

\section{Dual theory for neutral blackfolds}
\label{sec:3}

The AdS/CFT correspondence relates some classical results in a background spacetime to expectation value of quantum results at the boundary of that. For example the stress tensor which could be found in AAdS spacetimes could be taken as the expectation value of the stress tensor in a quantum conformal field theory. At the other hand Brown-York method does not work for all intrinsic metric in the reference spacetime and is not generally well defined. So in this section, we calculate the renormalized holographic stress energy tensor for the case of neutral blackfolds.
For every AAdS spacetime the expectation value of the dual CFT stress energy tensor is defined as:
 \begin{equation}
\label{eq:x}
<{T^{CFT}}_{ab}>=-\frac{2}{\sqrt{-g_{(0)}}}\frac{\delta S_{ren}}{\delta {g_{(0)}^{ab}}},
\end{equation}
$g_{(0)}$ is the metric on the boundary and $S_{ren}$ denotes the renormalized bulk action. In this case counterterms have been added to remove the volume divergences. We take advantages of Fefferman and Graham and obtain an asymptotic solution of Einstein's field equations with conformal structure. The corresponding solution is given by,
\begin{equation}
\label{eq:x}
ds^{2}=\emph{l}^{2}\left(\frac{d\rho ^{2}}{4\rho ^2}+\frac{1}{\rho}g_{ab}(x,\rho)dx^{a}dx^{b}\right),
\end{equation}
where $a,b= 1,...,D-1$  ,    $d=D-1$. And

\begin{equation}
\label{eq:x}
g(x,\rho)=g_{(0)}+...+\rho ^\frac{d}{2}g_{(d)}+h_{(d)}\rho ^\frac{d}{2} \log\rho+...,
\end{equation}
the coefficient $h_{(d)}$ is present only when $d$ is even. We will follow along the same line done in \cite{dd} and determine the explicit form of the dual renormalized holographic stress tensor,

\begin{equation}
\label{eq:x}
<{T^{CFT}}_{ab}>= \frac{d\emph{l}}{16\pi G_{N}}({{g_{(d)}}_{ab}}+{X^{(d)}}_{ab}),
\end{equation}
where ${X^{(d)}}_{ab}$ depends on the dimension. It's form for all even d is presented in ~\cite{dd}, and for odd d it vanishes,
\begin{equation}
\label{eq:x}
{X^{(2k+1)}}_{ab}=0.
\end{equation}
which reflects the fact that the holographic Weyl anomaly identically vanishes in even spacetime.
The Ricci-flat manifolds are Riemannian manifolds whose Ricci curvature vanishes.
We have checked that Ricci curvature of neutral black p-brane vanishes, and it is non zero for the case of charged black p-brane. So just the metric of (9) is Ricci-flat. There is a very nice connection between AdS black brane (which some of it's coordinates are compactified) and Ricci-flat spacetime. So one can find AAdS form of that metric.

 The neutral black p-brane in $D=n+p+3$ dimension has the form,
\begin{equation}
\label{eq:x}
 {ds^{2}}_{0}=-(1-\frac{{r_{0}}^{n}}{r^{n}})dt^{2}+{\sum _{i=1}}^{p}{dz_{i}}^{2}+{(1-\frac{{r_{0}}^{n}}{r^{n}})}^{-1}dr^{2} +r^{2}{d\Omega ^{2}}_{n+1},
\end{equation}
So the required components are given by,
\begin{equation}
\label{eq:x}
 {ds^{2}}_{p+2}=-(1-\frac{{r_{0}}^{n}}{r^{n}})\frac{\emph{l}^{2}}{r^{2}}dt^{2}+{\sum _{i=1}}^{p}\frac{\emph{l}^{2}}{r^{2}}{dz_{i}}^{2}+{(1-\frac{{r_{0}}^{n}}{r^{n}})}^{-1}\frac{\emph{l}^{2}}{r^{2}}dr^{2}\,,
\end{equation}
and
\begin{equation}
\label{eq:x}
 \phi(\rho,x)=(n+p+1)\ln(\frac{r}{\emph{l}}).\;\;\;\;\;\;\;\;\;\;\;\;\;\;\;\;\;\;\;\;\;\;\;\;\;\;\;\;\;\;\;\;\;\;\;\;\;\;\;\;\;\;\;\;\;\;\;\;\;\;\;\;\;\;\,
\end{equation}
 By changing $n \leftrightarrow -d$ , the AdS form of (43) will be planar  AdS black brane( with $\emph{l}=1 $ in $d+1$ dimension) is as follows,
\begin{equation}
\label{eq:x}
 {ds^{2}}_{\Lambda}=-\frac{1}{r^{2}}(1-\frac{{r}^{d}}{{r_{0}}^{d}})dt^{2}+ \frac{1}{r^{2}}({\sum _{i=1}}^{p}{dz_{i}}^{2}+d \vec{y}^{2})+\frac{1}{r^{2}(1-\frac{{r}^{d}}{{r_{0}}^{d}})}dr^{2},
\end{equation}
This metric is the AdS black brane and the Fefferman-Graham coordinate (39) is the best form for this theory to find the dual stress tensor. In order to do that, we must extract the relation between $r$ in (46) and $\rho$ in (39). So we have,
\begin{equation}
\label{eq:x}
\frac{1}{r^{2}(1-\frac{{r}^{d}}{{r_{0}}^{d}})}dr^{2}=\frac{d\rho ^{2}}{4\rho ^2},
\end{equation}
then,
\begin{equation}
\label{eq:x}
(\frac{r}{r_{0}})^{d}=\frac{4\rho ^{\frac{d}{2}}}{(\rho ^{\frac{d}{2}}+1)^{2}},
\end{equation}
The corresponding two important functions will be,
\begin{equation}
\label{eq:x}
\frac{1}{r^{2}}=\frac{1}{{r}_{0}^{2}4^{\frac{2}{d}}\rho}(\rho ^{\frac{d}{2}}+1)^{\frac{4}{d}},
\end{equation}

\begin{equation}
\label{eq:x}
\frac{1}{r^{2}}(1-\frac{{r}^{d}}{{r_{0}}^{d}})=\frac{1}{{r}_{0}^{2}4^{\frac{2}{d}}\rho}(\rho ^{\frac{d}{2}}+1)^{\frac{4}{d}-2}(\rho ^d -2\rho ^{\frac{d}{2}}+1),
\end{equation}
Consequently the AdS black brane in Fefferman-Graham coordinate take the following form,

\begin{equation}
\label{eq:x}
 {ds^{2}}_{\Lambda}=-\frac{1}{{r}_{0}^{2}4^{\frac{2}{d}}\rho}(\rho ^{\frac{d}{2}}+1)^{\frac{4}{d}-2}(\rho ^d -2\rho ^{\frac{d}{2}}+1)dt^{2}+ \frac{1}{{r}_{0}^{2}4^{\frac{2}{d}}\rho}(\rho ^{\frac{d}{2}}+1)^{\frac{4}{d}}[{dz_{i}}^{2}+d \vec{y}^{2}]+\frac{d\rho ^{2}}{4\rho ^2},
\end{equation}
In the case of $D=10 \;\;\rightarrow\;\; d=9$ , near the boundary $(\rho=0)$ the asymptotically AdS metric can be brought into the following geometry ,
\begin{equation}
\label{eq:x}
{ds^{2}}_{\Lambda}=-\frac{1}{{r}_{0}^{2}4^{\frac{2}{9}}\rho}(1-\frac{32}{9}\rho ^{\frac{9}{2}}+\frac{494}{81}\rho^{9}+...)dt^{2}+\frac{1}{{r}_{0}^{2}4^{\frac{2}{9}}\rho}(1+\frac{4}{9}\rho ^{\frac{9}{2}}-\frac{10}{81}\rho^{9}+...)[{dz_{i}}^{2}+d \vec{y}^{2}]+\frac{d\rho ^{2}}{4\rho ^2}
\end{equation}
By using the relation (40) one can obtain the explicit form of the metric coefficients which are given by,
\begin{equation}
\label{eq:x}
g_{(0)}=\frac{1}{{r}_{0}^{2}4^{\frac{2}{9}}}diag(-1,1,...,1),
\end{equation}

\begin{equation}
\label{eq:x}
g_{(9)}=\frac{1}{{r}_{0}^{2}4^{\frac{2}{9}}}diag(\frac{32}{9},\frac{4}{9},...,\frac{4}{9}),
\end{equation}
Finally the dual stress energy tensor in terms of relation (41) is

\begin{equation}
\label{eq:x}
<{T^{CFT}}_{ab}>= \frac{9}{16\pi G_{N}}({{g_{(9)}}_{ab}}+{X^{(9)}}_{ab})=\frac{9}{16\pi G_{N}}{g_{(9)}}_{ab}\;\;.
\end{equation}
 This stress tensor can be checked formally as a conserved and traceless quantity. It is traceless because the conformal anomaly evaluated for global AdS vanishes. If the metric (52) be locally boosted, it takes the form,
 \begin{equation}
\label{eq:x}
ds^{2}=-\frac{1}{{r}_{0}^{2}4^{\frac{2}{9}}\rho}(4\rho ^{\frac{9}{2}}-\frac{504}{81}\rho^{9}+...)u_{a}u_{b}dx^{a}dx^{b}+
\frac{1}{{r}_{0}^{2}4^{\frac{2}{9}}\rho}(1+\frac{4}{9}\rho ^{\frac{9}{2}}-\frac{10}{81}\rho^{9}+...)[\eta_{ab}dx^{a}dx^{b}+d \vec{y}^{2}]+
\frac{d\rho ^{2}}{4\rho ^2}.
\end{equation}
Then,
\begin{equation}
\label{eq:x}
{g_{(9)}}_{ab}=\frac{1}{{r}_{0}^{2}4^{\frac{2}{9}}}(4u_{a}u_{b},\frac{4}{9}\eta_{ab},...,\frac{4}{9}\eta_{ab}),
\end{equation}
So the renormalized holographic stress tensor takes the following general expression,
\begin{equation}
\label{eq:x}
<{T^{CFT}}_{ab}>= \frac{9}{16\pi G_{N}}[\frac{1}{{r}_{0}^{2}4^{\frac{2}{9}}}(4u_{a}u_{b}+\frac{4}{9}\eta_{ab})]\;\;.
\end{equation}
The boundary metric in 9-dimensional spacetime is conformally flat and the extracted stress energy tensor takes the ideal conformal fluid form . Comparison of (58) by (12) concludes the pressure and energy density of this blackfold
 such that,
 \begin{equation}
\label{eq:x}
p=\frac{1}{4^{\frac{11}{9}}\pi G_{N}r_{0}^{2}}, \;\; \;\;\;\;\;\;\;\;\; \varepsilon=\frac{8}{4^{\frac{11}{9}}\pi G_{N}r_{0}^{2}}\; . \;\;\;\;\;\;\;\;\;\;\;\;\;\;\
\end{equation}
It also proves that the resulted stress tensor (which determine the boundary fluid and comes from the duality of gravity and fluid)
is conserved and traceless.

For non AAdS  spacetime we can not use the represented method, because the holographic principle is not very well understood
for them. Recently some works on holographic connection for the asymptotically flat spacetimes have been done. Also it can be a good practice to extend the results of this paper to all higher dimensional black holes such as Myerse-Perry back holes and black Saturn. Using some mathematical techniques the conditions and constraints for holding such a holography could be extracted.

\section{Conclusion and outlook }
\label{sec:4}
In the covariant theory, the local energy momentum density of the gravitational field is not prevalent. Instead, a quasilocal stress tensor can be defined on the boundary of spacetime. But resulting stress tensor may diverge when the boundary goes to infinity. In order to solve this difficulty one can add an appropriate boundary term to the action  which is completely relevant.

Brown and York proposes a method to remove divergences. But it does not work for all intrinsic metric in the reference spacetime, so the Brown-York procedure is not generally well defined. Fortunately in AAdS spacetimes one can overcome to this problem. As gravitational action of the bulk corresponds to the (quantum effective) action of a conformal field theory on the AdS boundary, the expectation value of the stress tensor in the CFT solves the problem.

So this information gives us motivation to work the neutral blackfold and it's staff. So in the case of neutral blackfold eq. (9), the metric is Ricci-flat. AAdS form of pointed corresponding metric for this reason we are going to continue discussion on the neutral blackfold. In this paper we considered the metric of charged and neutral blackfold and obtained the boundary stress tensor. As the induced Brown-York stress tensor is conformally flat, the holographic stress tensor should be conformal to Brown-York stress energy tensor. For the case of neutral blackfold which it's metric is Ricci-flat, by applying the AdS/Ricci-flat correspondence, we found the corresponding AAdS form for it. Then for $D=10$ the renormalized holographic boundary stress tensor extracted. As we expected, the results show that the corresponding stress tensor are conserved and traceless. In that case there is not any conformal anomaly for the pointed theory. Also it takes the form of ideal fluid with determined energy density and pressure. \\

\end{document}